\documentclass[letter,12pt]{article}

\usepackage{graphicx}
\usepackage{color}
\usepackage{verbatim}
\usepackage[top=20mm, bottom=20mm, left=20mm, right=20mm]{geometry}
\definecolor{myblue}{rgb}{.0,.1,.5}
\definecolor{mygrey}{rgb}{.4,.4,.4}
\definecolor{bordeaux}{rgb}{.6,.09,.17}
\definecolor{burgundy}{rgb}{.62,.02,.03}

\def\Ca{{\rm Ca}^{2+}}

\def\Na{{\rm Na}^{+}}

\def\ncx{$\Na/\Ca$ exchange}

\def\iptr{IP$_3$R}

\def\bfig{\begin{figure}}
\def\endfig{\end{figure}}

\def\beq{\begin{equation}}
\def\endeq{\end{equation}}
\def\bt{\begin{table}}
\def\endt{\end{table}}
\def\bfig{\begin{figure}}
\def\endfig{\end{figure}}

\long\def\symbolfootnote[#1]#2{\begingroup%
\def\thefootnote{\fnsymbol{footnote}}\footnote[#1]{#2}\endgroup}

\begin{document}
%\pagenumbering{Roman}
\bibliographystyle{unsrt}
\title{Gold particle quantification in immuno-gold labelled sections
for transmission electron microscopy}
\author{Jeremy G. Hoskins, Nicola Fameli, Cornelis van Breemen\\
\emph{Department of Anesthesiology, Pharmacology,
and Therapeutics,} \\ \emph{The University of British
Columbia,} \\ \emph{Vancouver, British Columbia, Canada}}

\maketitle
\begin{abstract}
We briefly outline an algorithm for accurate quantification of specific
binding of gold particles to fixed biological tissue samples prepared
for immuno-transmission electron microscopy (TEM). The algorithm is based on
existing protocols for rational accounting of colloidal gold particles
used in secondary antibodies for immuno-gold labeling.
\end{abstract}
\section{Introduction}
Vascular smooth muscle cells in intact tissue are the primary focus of
the work done in our laboratory. Different 
TEM samples from the same tissue batch are
usually prepared both for ultrastructural imaging and for immuno-gold labeling.
The former yields, among other features, a picture of the distribution of
membrane-bound components such as mitochondria, SR, lysosomes, Golgi
apparatus and plasma membrane. We then perform immuno-gold labeling of
ultrathin (50 to 80 nm) sections with primary antibodies typically
targeting relevant transport or membrane proteins. Specifically, our
laboratory is interested in various isoforms of 
sarco/endoplasmic reticulum ATPases (SERCA),
lysosome associate membrane proteins (LAMP), \ncx\ proteins, 
ryanodine receptors
(RyR), and \iptr.
\section{Gold Detection Algorithm}
The electron micrographs produced during the imaging 
of tissue samples prepared for immunogold
labeling typically contain faint grey-scale images of the cells,
organelles and nuclei.
Gold nanoparticles used for labeling appear
round and very dark, with a fixed
size, visible only
at a magnification of 30000$\times$ or more. There may be other dark
regions, usually oddly shaped, lighter than the gold, and of varying
in size, that are most likely
artifacts introduced in the fixing, or staining processes. 
Using the scripting language Python, we wrote a software algorithm for
automatic detection of gold nanoparticles. It consists of 
three main steps, addressed in the following subsections: contrast
filtering, spot detection, size filtering, and shape filtering.
\subsection{Constrast filter}
The contrast filter, as implemented in the Python script, is designed
to sort the grayscale
pixels into two categories: those that are potentially gold and those
unlikely to be gold.
This is based upon the observation that gold particles are
significantly darker than the rest
of the cell. Note that for the electron micrographs, higher density
particles corresponds to
a lower intensity value of the associated image pixel. Gold particles,
being dense, therefore
appear darker than the rest of the image.

The contrast filter uses an arbitrary intensity cut-off,
anywhere from 10 to 40 out of 255, by which it is possible to retain a
large percentage of
the gold pixels, whilst keeping little of the background material. The
higher this intensity
cut-off is, the greater the percentage of gold particles that will be
detected. Increasing this
value does, however, also increase the amount of non-gold which passes
through the filter,
which in turn increases the chance of a false-positive. 
A right balance is struck by appropriate algorithm validation against
reference images. Though the implemented version
converts the image into this binary array, it would also be possible to
raise the cut-off and
store in the array a probability of the specified pixel being part of a
gold particle. Thus, a
pixel close to the cut-off would still be stored as a potential gold
location, but would have
less weight in the following algorithms than a pixel with a lower
intensity value.
\subsection{Spot detection}
The spot detection portion of the algorithm detects dense areas and
separates them into
different regions, called spots. Two pixels lie in distinct spots if
there exists no path joining
the two pixels, which lies entirely within the high-density regions.
Here a path represents
a series of pixels which share a common edge. The algorithm works by
rastering over the
pixel grid generated from the contrast filter, and performing a
flood-fill on every dark pixel
it encounters. The flood-fill stores all the location of all the pixels
in the same spot as
the original pixel. Then, the region is turned white, to avoid
repeating the same pixel
in two distinct spots. A flood-fill algorithm is a recursive algorithm
for determining all
pixels which lie in a given boundary, or, as in this case, all pixels
in a simply-connected
region, which have the same colour. It grows by checking the neighbours
of all the pixels
in the current region, and, if they match the criteria, adding them to
the new region. This
process continues until no new additions take place between successive
iterations.
\subsection{Size filter}
Gold particles are, by their fabrication, of constant size ($\pm 0.56$
nm) \cite{Yang2008}. 
When trying to determine
which of the
dark spots are gold, this criterion is useful for providing a
preliminary filter. 
In the images, apart from the problems with pixelation, 
all gold particles occupy roughly the same
number of pixels.
Thus, introducing a band pass filter on spot size guarantees that only
particles close in
size to gold are considered as such by the algorithm. 
Single pixels, which are ’round’ like gold, and
large, round artifacts are eliminated at this stage in the algorithm.
Since each spot is
stored as a list of locations, one can easily implement this filter by
simply comparing the
size of this list to the expected parameters. Provided images have
similar magnification
and contrast, choosing these parameters need only be done once.
\subsection{Shape filter}
The shape filter utilizes another of the key characteristics of the
gold particles: their
projection onto a plane, such as what occurs when taking an electron
micrograph, is a
circle. In the Python script, for each spot the sum of the squared
distance of each location
from the spot centroid is calculated. Dividing this by the area squared
gives a quantity
which will be at a minimum for circles, where the value will be 1/2.
Alternatively, one could use instead 
the eccentricity, $e$. Determining the largest and smallest distance of an
edge pixel from the centroid, $r_{\rm max}$ and $r_{\rm min}$ 
respectively, it can be
calculated using:
\begin{equation}
e={r_{\rm max}-r_{\rm min}\over r_{\rm max}+r_{\rm min}}
\end{equation}
For a circular object, this value should be zero, with higher values
corresponding to regions
which have less radial symmetry. Using either method, once the quantity
for comparison  
has been calculated, a low pass filter may be employed to eliminate any
non-circular
objects. 

The right panel of Figure \ref{figure_1} shows the result of the algorithm
with colours inverted for clarity.
\bfig[tb]\centering
\includegraphics[scale=.6]{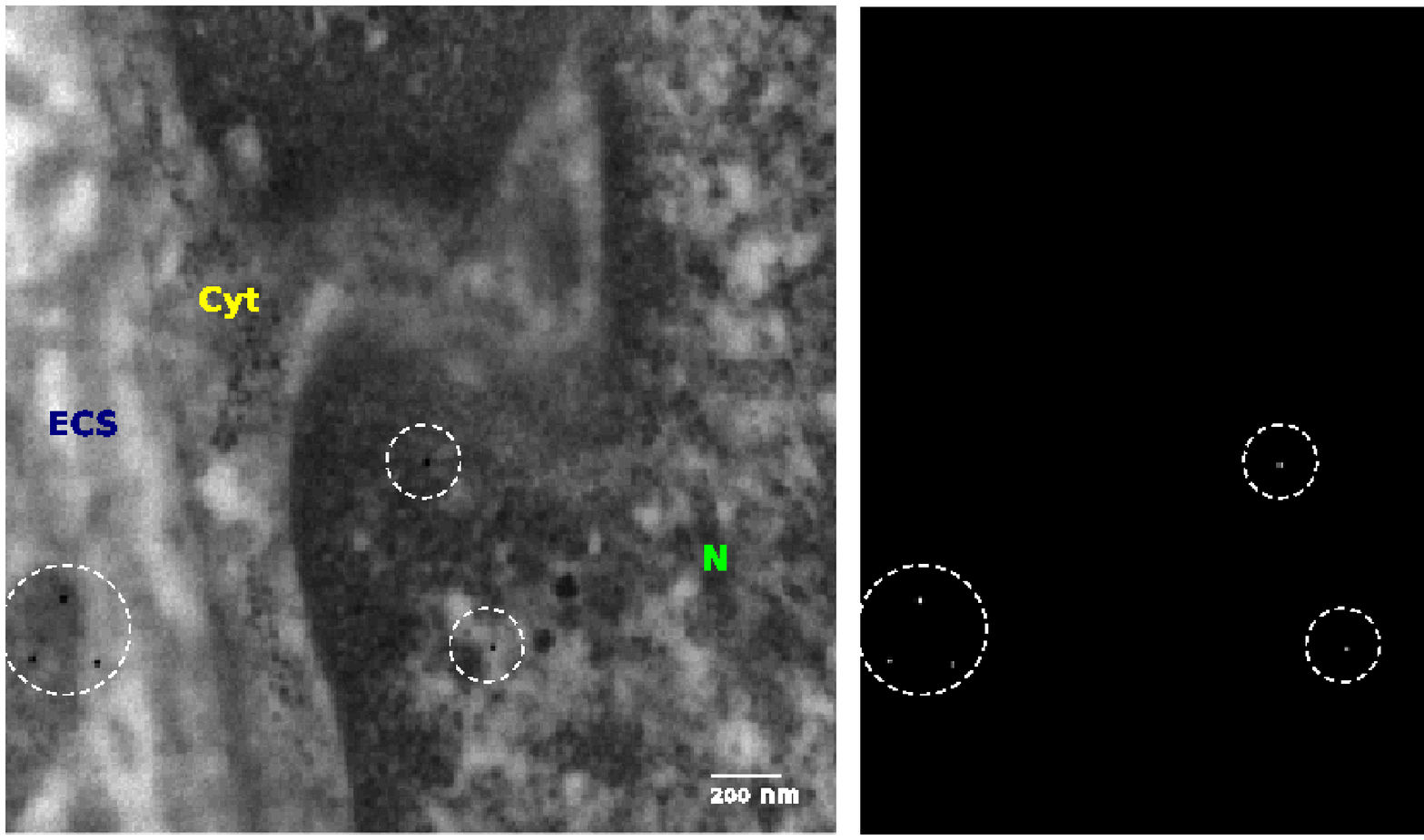}
\caption{\small
\textbf{Left:} Electron micrograph of rat pulmonary artery stained for
SERCA1 (gold particles inside
dashed circles). \textbf{Right:}
Result of gold detection algorithm applied to the image on the left.
Gold is represented
as white. \textbf{ECS}=extracellular space, \textbf{Cyt}=cytoplasm, 
\textbf{N}=nucleus.}\label{figure_1}
\endfig
\section{Gold Preferential Localization Analysis}
\subsection{Localization ratio}
The utility of the gold detection algorithm could be extended to include 
unbiased region and membrane localization analysis. 
The specificity of localization to a region M can be determined by the ratio:
\begin{equation}
R={G_M/G_C\over A_M/A_C}
\end{equation}
where $G_M$ and $G_C$ are the number of gold particles in the region and cell
respectively, and
$A_M$ and $A_C$ are their areas \cite{Mayhew2003b}. 
A value of 1 corresponds to a near random
distribution. A
ratio higher than 1 suggests localization whilst a ratio less than 1
signifies an absence of
that particle in the specified region. The farther the deviation from
1, the greater is the
extent of localization to, or absence from, the region. Assuming the
gold, if unlocalized,
would follow a binomial distribution with probability $p = A_M/A_C$ ,
i. e., a uniform probability
distribution over the area of the cell, the standard deviation expected
would be $\sigma = \sqrt{np(1-p)}=\sqrt{G_C {A_M\over A_C}(1-{A_M\over A_C})}$. 
This can be used to test the likelihood
that a given
particle distribution is the product of random placement.
\subsection{Membrane and region selection}
By using the counting algorithm described above, one can count the
numbers of particles in specified regions and calculate the value of 
$R$ for various regions or membranes of interest, as identified 
in a computer readable
format (we used the software package `inkscape').
This is carried 
by outlining the region, or membrane, of interest in a closed path. 
For the membrane,
the `Halo' function in inkscape is then used, 
%with the pixel distance set such that
%the distance from
%the membrane corresponds to 50 nm. 
A custom Python script was written
which exports
these closed paths as lists of node locations. The entire cell is also
outlined, to give the
whole cell area for comparison.
\subsection{Area Calculation}
The area of the closed region defined by a boundary path is determined
by first choosing
either the $x-$ or $y-$direction \cite{Goodchild1990}. 
The program then loops through each
pair of successive
nodes and calculates the signed area of the trapezoid relative to that
axis. Thus, if the
points happen to be in reverse order relative to that axis, the
trapezoid will have a negative
area. The areas of all the trapezoids are summed and the resultant
value is, up to a sign,
the total area of the desired region. In order to deal with the area 
%50nm 
a given distance from a specified
line or point, one could instead use a flood-fill algorithm which tests
the distance from the
line for each prospective point.
\bfig[tb]\centering
\includegraphics[scale=.4]{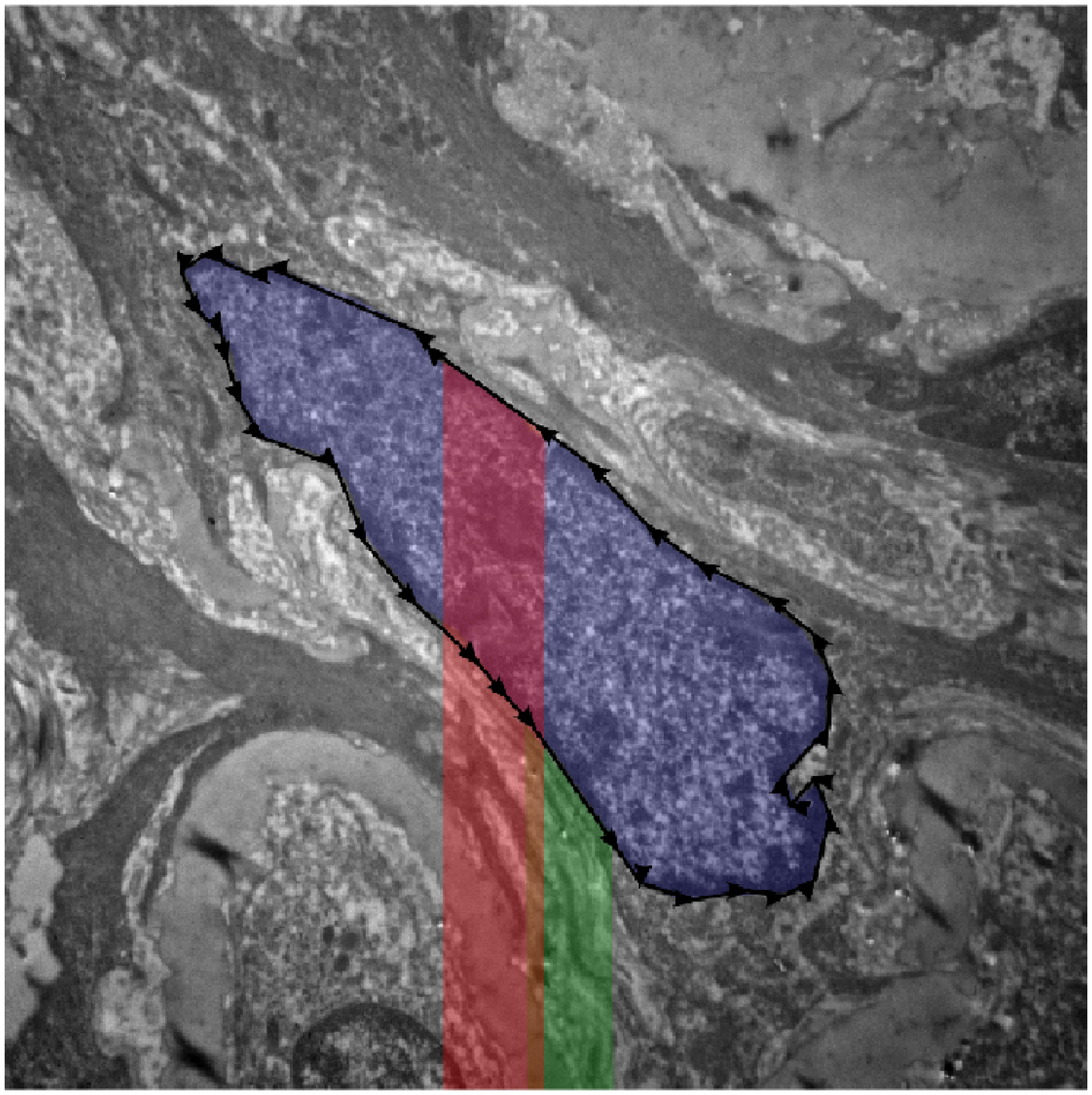}
\caption{\small A stylized depiction of the operation of the area
algorithm. The region shaded
in blue is the area to be measured. The red trapezoid, as the arrows
suggest, is one with
a negative area contribution whilst the green trapezoid yields a
positive contribution.}\label{figure_2}
\endfig
\subsection{Whole Cell Analysis}
If the image of the individual parts of the cell can be stitched
together, then the membranes or 
regions can be left in place. Then, when the gold detection
algorithm runs, the
software can determine whether or not each gold particle is inside the
region. The ratio
can then be calculated for the entire cell. If stitching is not
possible, then one can take a
series of pictures either of the whole cell or randomly selected
regions. In this case, one
must then face the attendant problem of ensuring the images are a truly
random sample,
with a sample size large enough to ensure that enough of the desired
membrane, or region
is present. The distance of the gold can be determined either manually
or electronically by
seeing, for each picture, if any region is present, and where the gold
particles are distributed
relative to that region.
\bibliography{Au_detection}
\end{document}